\documentclass[epjST]{svjour}
\usepackage{graphicx}
\usepackage{amssymb}
\usepackage{amsmath}
\usepackage{bm}
\usepackage{xcolor}
\usepackage{float}

\begin{document}
\title{Tipping induced by multiplexing on two layer networks}%
\author{Umesh Kumar Verma \fnmsep\thanks{\email{umeshvermaphy@gmail.com}} and G. Ambika \fnmsep\thanks{\email{g.ambika@iisertirupati.ac.in}}}
\institute{Indian Institute of Science Education and Research(IISER) Tirupati, Tirupati, 517507, India}

\date{\today}

\begin{abstract}
{We report the study of sudden transitions or tipping induced in a collection of systems due to multiplexing with another network of systems. The emergent dynamics of oscillators on one layer can undergo a sudden transition to steady state due to indirect coupling with a shared environment, mean field couplings and conjugate couplings among them. In all these cases, when multiplexed with another set of similar systems, the tipping phenomena are induced on the second layer also with a similar pattern of behavior. We consider van der Pol oscillator as nodal dynamics with various network topologies like scale-free and regular networks with local and nonlocal couplings. We also report how the coupling topology influences the nature of transitions on both layers, under multiplexing.}
\end{abstract}


\maketitle

\section{Introduction }

Recent studies on sudden transitions in the collective behavior of complex systems are quite widespread, including several systems such as epileptic seizures in the brain~\cite{R112} jamming in the Internet~\cite{R2}, and sudden shifts in climate~\cite{R3} and ecological systems~\cite{R4,R5}.  In climate science, such sudden transitions are termed as {\it tipping points}, and in ecology, as {\it regime shifts}~\cite{R6}.  A few well-studied cases are, shutdown of the thermohaline ocean circulation, the transition from a wet to a dry Indian monsoon system~\cite{R7,R8}, the loss of sea ice in the Arctic~\cite{R9,R10}, changes in various ecosystems~\cite{R4,R11} such as coral reefs~\cite{R12,R13}, shallow lakes~\cite{R14},or vegetation patterns in semiarid areas~\cite{R15,R16} or in the Sahara~\cite{R17}.  Attempts to predict and prevent undesirable tipping, especially those related to climate changes, are important for humanity to achieve sustainable development.  

Such tipping phenomena are well studied in multi-stable systems~\cite{R18,R19,R20}, where qualitatively different states coexist for a given set of parameters.  In such systems tipping elements are nonlinear dynamical systems, where a small perturbation or change in parameter near a critical point can induce a qualitative change in the whole system as soon as a critical threshold or tipping point is crossed. However, in several cases, we come across interacting tipping elements like the climate system, interacting species in ecosystems, physiological systems, coupled patches of vegetation, disease spreading in human populations, and detuning in multiscale infrastructure systems~\cite{RR1,RR2}.  For such systems that are most often spread across space, the framework of complex networks is most useful to study the transitions in their collective dynamics. Each node in the network then becomes a tipping element, and the topology of their connections on the network decides the robustness or stability of dynamics. Recent studies indicate that clustering, high motif frequency, and spatial organization increase the vulnerability of such networks and can lead to the tipping of dynamics on the whole network~\cite{RR20}. 

The mechanisms underlying sudden transitions that happen when one or more parameters of the system reach critical values are identified as bifurcation induced(B-tipping) ~\cite{R21,R22}, noise induced(N-tipping)~\cite{R23}, when tipping happens in the presence of noise or rate induced (r-tipping)~\cite{R24}, when a change in one parameter at a certain rate leads to tipping. In this study, we introduce a new mechanism where tipping is induced by one layer on the other in a two-layer network of systems due to multiplexing between layers.

Recent studies indicate that, multilayer networks where the nodes are distributed in different layers offer a better model for studying the interactions and dynamics of real-world systems~\cite{RN3,RN4,RN5}.  Among them,  multiplex networks form a specific set where nodes are the same in different layers  and thus provide a natural way to model different dynamical interactions among the same set of dynamical systems.  From social media, economy, transportation systems to ecological and biological systems, such interdependencies exist among nodes and multiplex networks are the ideal framework for their study. In such cases, it is interesting to study how the processes or organization happening in one layer network drive the other in many significant ways, inducing transitions in them. Some of the studies in this direction, relevant in the present context, are congestion phenomena on multiplex transportation networks~\cite{RN6},  explosive synchronization induced by multiplexing between excitatory and inhibitory coupled layers~\cite{RN7}, a discontinuous transition in layers of consensus dynamics and information spreading in multiplexed social networks~\cite{RN8}, weak multiplexing induced chimeras in neural networks~\cite{RN9} etc. Transitions with multiple tipping points, as well as complex hysteresis loop, are reported in interacting networks with optimal, repairing strategy~\cite{RN10}.

In this work, we consider the multiplex network as the basic framework, with one layer having van der Pol oscillators coupled with various coupling schemes such as scale-free(S-F) and regular networks with nonlocal and global couplings. We study the occurrence of sudden transitions on the layer due to various types of couplings like indirect coupling with a shared environment, due to mean-field coupling and conjugate coupling. We find, near the critical transition point, both oscillatory state and steady-state coexist, and with an adiabatic change of coupling parameters, hysteresis is observed, indicating first-order transitions. When this layer is multiplexed with another set of similar systems, we observe that similar transitions are induced in them also. When the systems in the second layer are coupled with the same topology as the first layer, with S-F structures, the transitions induced in them are sudden but with no hysteresis. However, when the systems in the second layer have a different coupling scheme, the transitions in both layers follow a pattern that depends on the nature of couplings. Thus in systems where tipping is induced due to multiplexing, the heterogeneity in the coupling of the layers as well as the mechanism for tipping in the first layer decides the nature of transitions on both layers.

\section{Multiplex networks}
We start by constructing a two-layer network of nonlinear oscillators, such that oscillators in the first layer $L_1$ interact with oscillators in the second layer $L_2$ with multiplex like $i$ to $i$ coupling. The dynamics of $N$ nonlinear oscillators on the multiplex network thus, modeled is in general given by,

\begin{eqnarray}
 \dot{\bm{X}}_{i,1} & = \bm{F}(\bm{X}_i)+\kappa_1\sum_{j=1}^{N}A_{i,j}(x^1_{j,1}-x^1_{i,1})+\eta(x^1_{i,2}-x^1_{i,1}), \nonumber\\
 \dot{\bm{X}}_{i,2} &= \bm{F}(\bm{X}_i)+\kappa_2\sum_{j=1}^{N}B_{i,j}(x^1_{j,2}-x^1_{i,2})+\eta(x^1_{i,1}-x^1_{i,2}), 
\label{eq:eqmodel}
\end{eqnarray}
where $i=1,2,\ldots,N$. Here $\bm{X}_{i,l}= \begin{bmatrix} x^1_{i,l} & x^2_{i,l} & \ldots & x^m_{i,l} \end{bmatrix}^{T}$ represent $m$--dimensional state space of the oscillator whose intrinsic dynamics is given by 

$\bm{F}(\bm{X}_{i,l})=\begin{bmatrix} f^1_{i,l}(\bm{X}_{i,l}) &f^2_{i,l}(\bm{X}_{i,l})& \ldots & f^m_{i,l}(\bm{X}_{i,l}) \end{bmatrix}^{T}$, where $l=1,2$. $A_{i,j}$ and $ B_{i,j}$ are constant adjacency matrices having dimensions $N \times N$  and their elements depend on the specific coupling scheme chosen. The oscillators in the first layer interact with oscillators in the second layer with multiplex like $i$ to $i$ coupling with coupling strength $\eta$.  $\kappa_1$, and $\kappa_2$  represent the strengths of the coupling of the  oscillators in each layer $L_1$ and $L_2$ respectively.

In such multiplex networks, we study the phenomena of the sudden transitions from oscillatory state to amplitude death state and vice versa, which occur in one layer $L_1$ and get induced on the other layer $L_2$. We consider different network topologies for $L_1$ such as S-F network, nonlocal and global couplings, conjugate coupling and mean-field coupling to illustrate our results. With multiplexing, we consider cases when the systems in the second layer $L_2$ are uncoupled and coupled with the same topology and different topology as $L_1$. In all these cases, we present the nature of the tipping or sudden transitions with the adiabatic drift of parameters, along with the space time plots and the parameter planes to indicate various emergent dynamical states in the system.

To characterize the collective behavior of the coupled systems with the coupling strength $\lambda$, and to identify the tipping point or sudden transitions, we define an order parameter $A(\lambda)=\frac{a(\lambda)}{a(0)}$, where 
\begin{equation}
\begin{split}
  a(\lambda) = \frac{1}{N}\sum_{i=1}^N \left[\langle x_{i,max} \rangle_t - \langle x_{i,min}\rangle_t \right].
\end{split}
\label{eq:defae}
\end{equation}

The value of $A(\lambda)=0$ when all oscillators are in the stationary state of amplitude death, whereas when all oscillators are in the oscillatory state, the value of $A(\lambda)>0$. The order parameter $A(\lambda)$ is calculated for an adiabatic change of coupling strength both in forward and backward directions. We start by solving for the dynamics of coupled oscillators for random initial conditions, and $A(\lambda)$ is calculated, then the value of $\lambda$ is gradually increased from $\lambda_0$ to $\lambda_{max}$ with steps size $\delta \lambda=0.02$. At each stage, the final state for the earlier $\lambda$ is used as the initial condition for the next $\lambda$. The backward continuation is also performed in the same way by gradually decreasing the value of $\lambda$ from $\lambda_{max}$ to $\lambda_0$.

\section{Coupled van der Pol oscillators with shared environment}
We first consider a multiplex network with N identical van der Pol (VdP) oscillators on each layer. As one of the mechanisms for sudden transitions in dynamics to amplitude death, in this section, we consider a common shared environment interacting with systems in the first layer $L_1$ only. The oscillators of the second layer $L_2$ can be uncoupled or coupled with various coupling schemes. The dynamics of the multiplex network is, thus, given by,

\begin{eqnarray}
\dot{x}_{i,1}&=&y_{i,1}+ \frac{\lambda_1}{k_{i,1}}\sum_{j=1}^{N}A_{i,j}(x_{j,1}-x_{i,1})+\epsilon s +\eta (x_{i,2}-x_{i,1}) \nonumber\\
\dot{y}_{i,1}&=&b(1-x_{i,1}^2)y_{i,1}-x_{i,1} \nonumber\\
\dot{x}_{i,2}&=&y_{i,2}+\frac{\lambda_2}{k_{i,2}}\sum_{j=1}^{N}B_{i,j}(x_{j,2}-x_{i,2})+\eta(x_{i,1}-x_{i,2})\nonumber\\
\dot{y}_{i,2}&=&b(1-x_{i,2}^2)y_{i,2}-x_{i,2} \nonumber\\
\dot{s}&=&-\gamma s-\epsilon \bar{x}_{i,1}
\label{eq3}
\end{eqnarray}

where $b$ is the damping coefficient of each oscillator and $b > 0$, the system generates limit cycle oscillations. $(x_{i,1},y_{i,1})$ represent oscillators of $L_1$ and $(x_{i,2},y_{i,2})$ those of $L_2$, with $i=1,2....N$. The connections among oscillators in each layer are encoded in the adjacency matrices of the network, $A$ and $B$. The oscillators in $L_1$ interact with oscillators in $L_2$ with multiplex like $i$ to $i$ coupling with coupling strength $\eta$. The oscillators in layer $L_1$ and $L_2$ interact among them with coupling strength $\lambda_1$ and $\lambda_2$ respectively. Here $k_{i,1}$ and $k_{i,2}$ are the average degree of $i^{th}$ oscillators  corresponding to each of the layers $L_1$ and $L_2$ respectively. The variable $s$ represents the dynamics of the common shared environment, modeled by an over-damped oscillator with decay rate $\gamma>0$ in the absence of the coupling~\cite{RR28,RR29,RR30}. The environment gets average negative feedback from all systems, while it gives positive feedback to the systems, with a coupling strength $\epsilon$. The environment, depending on the context, can represent common particle species that can freely diffuse in the system and allow individual oscillators to communicate with each other or a shared homogeneous medium ~\cite{RR31,RR32,RR33}

\begin{figure}
\includegraphics[width=0.75\textwidth]{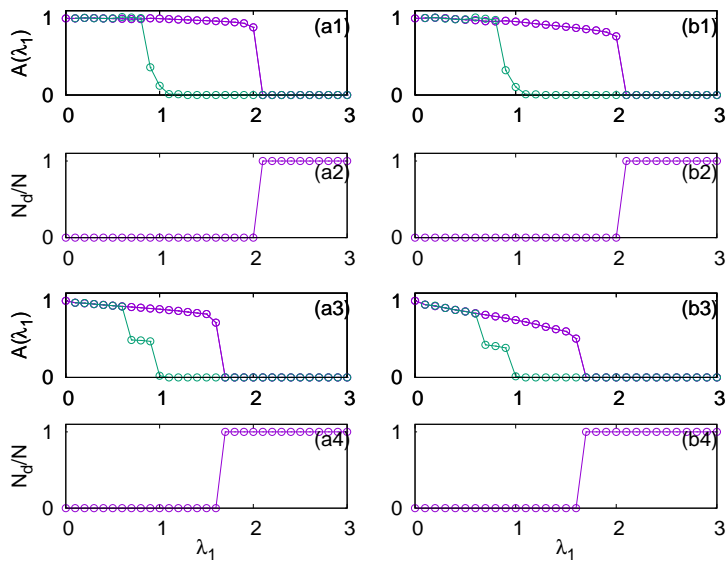}
\caption{Transitions from oscillatory state to death state in multiplex network of coupled van der Pol oscillators. The layer $L_1$ (left panel) has S-F topology of coupling, along with a shared environment and layer $L_2$ (right panel) form a set of uncoupled systems, with $N$ oscillators in each. {\bf a1, b1} The order parameter, $A(\lambda_1)$, under adiabatic variation of the coupling strength $\lambda_1$, in forward(red) and backward(green) directions with $<k_1>=6$; {\bf a2, b2} The ratio of oscillators in death state to the total number, $Nd/N$, plotted in forward continuation with $< k1 >= 6$; {\bf a3, b3} The order parameter during transition with $<k_1>=20$ and {\bf a4, b4} $Nd/N$, with $< k1 >= 20$. The other parameters are $b=1$, $\epsilon=1.5$, $\eta=1$, $\lambda_2=0.0$, and  $N=500$.  The tipping occurring on $L_1$ induces a similar pattern of transitions on $L_2$ even when the oscillators of $L_2$ are uncoupled. The tipping point gets shifted with change in $< k1 >$}
\label{f1}   
\end{figure}

\subsection{Multiplex network with scale-free topology}
We begin with a S-F topology for both layers of the network and first study the emergent dynamics on each layer in the absence of multiplexing with $\eta=0$ in Eqn.~\ref{eq3}. Here, we define  average degree of each layer network, as $<k_1>=<k_{i,1}>$ and $<k_2>=<k_{i,2}>$. We fix $\epsilon=1.5$, $b=1.0$, $<k_1>=6$, and $N=500$ and adiabatically vary $\lambda_1$ in forward and backward directions for layer $L_1$. In the forward direction, we observe a sudden change in order parameter $A(\lambda_1)$ from nonzero to zero, and in the backward direction, also a sudden change in $A(\lambda_1)$ from zero to nonzero. These changes indicate a sudden transition from oscillatory state to death state and vice versa in both forward and backward directions. The transition from oscillatory state to death occurs due to inverse Hopf bifurcation. Depending on the parameters of coupling, the system can stabilize at a coupling-dependent Inhomogeneous Steady State(IHSS), where the coupled oscillators form clusters that settle to either $x_i \approx 1$ or $x_i\approx -1$. The coupled system can also stabilize at coupling dependent Homogeneous Steady State (HSS)where all oscillators stabilize at the same steady state. We also observe that in the absence of coupling with the environment, there is no transition to death in layer $L_2$. 

Now we consider multiplexing both the layers, where the oscillators in $L_1$ are connected with S-F topology. As a limiting case, we first consider the oscillators of $L_2$ as uncoupled (i.e., $\lambda_2=0$), and both layers are coupled to each other via $i$ to $i$ connections with coupling strength $\eta$. We set $b=1$, $\epsilon=1.5$, $\eta=1$, and $<k_1>=6$ and vary $\lambda_1$ adiabatically. In forward direction, we observe a sudden transition from the oscillatory state to death state in both layers $L_1$ and $L_2$, which is shown in Fig.~\ref{f1}a1, b1. Similarly, in backward direction also, we observe a sudden transition from death state to oscillatory state, in $L_1$ and $L_2$. The ratio of oscillators in death state to the total number, $N_d/N$, for both layers in the case of forward direction is plotted in Fig.~\ref{f1}a2, b2. For $<k_1>=20$, the forward and backward variations for both layers is plotted in Fig.~\ref{f1}a3, b3 with the corresponding $N_d/N$, in Fig.~\ref{f1}a4, b4.

\begin{figure}
\includegraphics[width=0.7\textwidth]{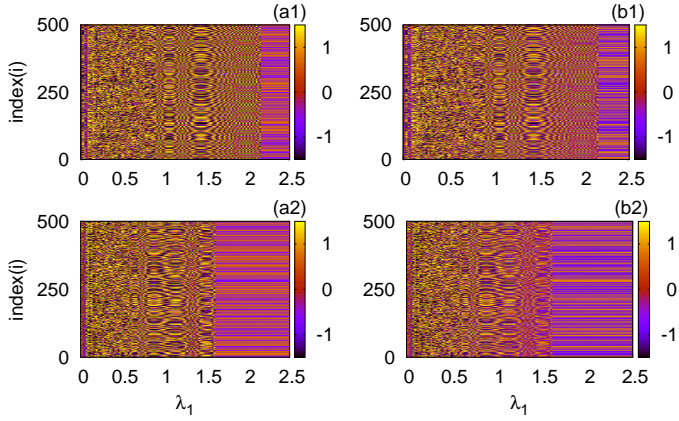}
\caption{Transitions from oscillatory state to inhomogeneous steady state with increasing intra layer coupling strength. Variation of variables $x_i (i = 1,2,...,500)$ for $L_1$  (left panel) and $L_2$ (Right panel) shown in $(\lambda_1,i$) plane showing transitions from oscillations to IHSS: {\bf a1, b1} $<k_1>=6$, and {\bf a2, b2}  $<k_1>=20$.  The other parameters are as in Fig.1. With multiplexing, the type of transition on $L_1$ is induced on the set of uncoupled systems forming $L_2$. }
\label{f2}   
\end{figure}

We note that the tipping phenomena occurring on $L_1$ induce a similar pattern of transitions on $L_2$ even when the oscillators of $L_2$ are uncoupled.  To display the consequent changes in dynamics, we plot the average $x_i (i = 1,2,...,500)$ for both layers $L_1$ and $L_2$ in Fig.~\ref{f2}a1, b1 for $<k_1>=6$ and Fig.~\ref{f2}a2, b2 for $<k_1>=20$, where the average is taken over a long time after transients. With increase in the average degree, the tipping point shifts to smaller values of $\lambda_1$.  At the tipping point, all oscillators show a transition from oscillatory state to death state, and in the death state, they display a state of Inhomogeneous Steady States(IHSS). In this case, near the transition point, there is bi-stability where oscillatory state and IHSS state coexist, leading to transition with hysteresis. As mentioned, these transitions can happen in single layer of network. With multiplexing, the additional transition is layer to layer in-phase synchronization or pairwise in-phase synchronization of nodes between layers, which triggers tipping in both layers. We observe, this can happen for intralayer coupling for which the systems are not synchronized within each layer before tipping.

\begin{figure}
\includegraphics[width=0.5\textwidth]{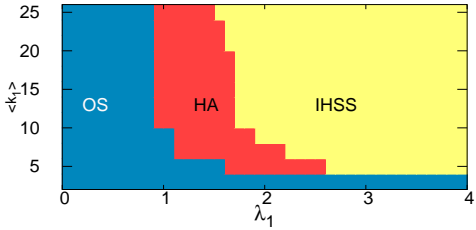}
\caption{Parameter plane showing regions of different emergent states for multiplex network with S-F topology in $L_1$ . The emergent states depicted are oscillatory state(OS), hysteresis area(HA) and inhomogeneous steady state(IHSS)for $\epsilon=1.5$, $b=1.0$, $\eta=1.0$, $\lambda_2=0$ and $N=100$. }  
 \label{f3}  
\end{figure}

We show the parameter plane $(\lambda_1, <k_1>)$ for layer $L_1$, in Fig.~\ref{f3}. We find the possible emergent states are oscillatory state(OS), hysteresis area(HA) (where oscillatory state and death state coexist), and inhomogeneous steady state(IHSS). The transitions points are found to shift to the left for a higher values of $<k_1>$. Similar states are observed for the emergent dynamics of systems in layer $L_2$ also indicating the effect of multiplexing.

\begin{figure}
\includegraphics[width=0.65\textwidth]{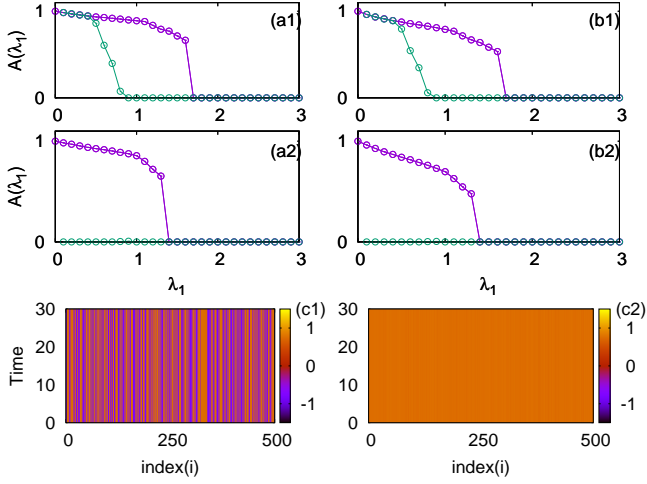}
\caption{ Transitions from oscillatory state to steady state for $N$ coupled VdP oscillators in $L_1$ (left panel) and $L_2$ (right panel). {\bf a1,b1} The variation of $A(\lambda_1)$, with the coupling strength $\lambda_1$ for $<k_1>=6$, and $<k_2>=4$ and {\bf a2, b2}  for $<k_1>=20$, and $<k_2>=4$. {\bf c1, c2} The spatio temporal plot in steady state regime  for layer L1 {\bf c1} at $\lambda_1=2$: for $<k_1>=6$,  $<k_2>=4$  and {\bf c2} at $\lambda_1=1.8$: for $<k_1>=20$, and $<k_2>=4$. Here $b=1$, $\epsilon=1.5$, $\lambda_2=0.20$, $\eta=1.0$, and $N=500$. With increasing $<k_1>$ of layer $L_1$ , the transition occurs with no hysteresis.  }  
\label{f4}   
\end{figure}

We extend the study to the case when oscillators of layer $L_2$ are also coupled with a S-F topology and multiplexed with the first layer. Here we first set $<k_1>=6$,  $<k_2>=4$, $\lambda_2=0.2$, and $\epsilon=1.5$, and study variation of the order parameters $A(\lambda_1)$ for both the layers, which are shown in Fig.~\ref{f4}a1, b1 respectively. We find in the forward direction both layers show sudden transitions from oscillatory state to death state. However, in the backward direction, the transition is not sudden, but both layers still show hysteresis indicating coexistence of  both oscillatory and death states. In this case, after the transition, the systems stabilize to IHSS on both layers (see Fig.~\ref{f4}c1). When we increase $<k_1>$ to $20$, the transition is sudden in the forward direction, but there is no transition from the death state to oscillatory state in the backward direction, as is clear from Fig.~\ref{f4}a2, b2. In this case, after the transition the systems stabilize to Homogeneous Steady State(HSS)on both layers(Fig.~\ref{f4}c2). Thus the nature of transition depends on the heterogeneity in the nature of connections in $L_1$.

\subsection{Multiplex network with regular and nonlocal connections}

\begin{figure}
\includegraphics[width=0.75\textwidth]{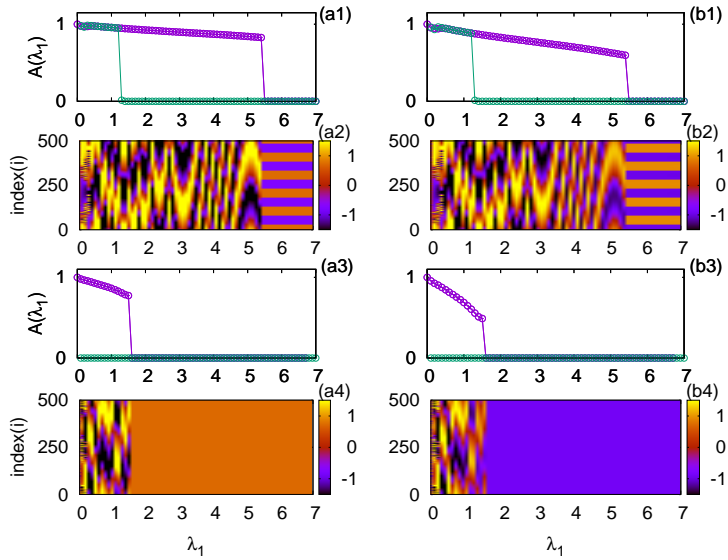}
\caption{Transitions from oscillatory state to death state in multiplex network with regular and nonlocal connections. The layer $L_1$ (left panel) with regular nonlocal connections and layer $L_2$ (right panel) has uncoupled systems. {\bf a1, b1} The order parameter, $A(\lambda_1)$, under the variation of the coupling strength $\lambda_1$ for  $R_1=0.08$ and {\bf a3, b3} for $R_1=0.2$. {\bf a2, b2} The variation of variables $x_i (i = 1,2,...,500)$  with increasing  $(\lambda_1,i)$ for  $R_1=0.08$ and {\bf a4, b4} for $R_1=0.2$.  We fix $b=1$, $\epsilon=1.5$, $\lambda_2=0.0$, and  $N=500$. For $R_1=0.08$, the steady state is IHSS with hysteresis, while for $R_1=0.2$, it is HSS with no hysteresis.}
\label{f5}   
\end{figure}

The dynamics of the multiplex network of N identical VdP oscillators with regular and nonlocal connections is given by,

\begin{eqnarray}
\dot{x}_{i,1}&=&y_{i,1}+ \frac{\lambda_1}{2P_1}\sum_{j=i-P_1}^{i+P_1}(x_{j,1}-x_{i,1})+\epsilon s +\eta (x_{i,2}-x_{i,1}) \nonumber\\
\dot{y}_{i,1}&=&b(1-x_{i,1}^2)y_{i,1}-x_{i,1} \nonumber\\
\dot{x}_{i,2}&=&y_{i,2}+\frac{\lambda_2}{2P_2}\sum_{j=i-P_2}^{i+P_2}(x_{j,2}-x_{i,2})+\eta(x_{i,1}-x_{i,2})\nonumber\\
\dot{y}_{i,2}&=&b(1-x_{i,2}^2)y_{i,2}-x_{i,2} \nonumber\\
\dot{s}&=&-\gamma s-\epsilon \bar{x}_{i,1}
\end{eqnarray}

where $P_1$ and $P_2 \in \{1, N/2\}$, correspond to the number of nearest neighbors in each direction on each layer and corresponding coupling radii are $R_1=\frac{P_1}{2N}$ and $R_2=\frac{P_2}{2N}$ respectively. For single network with regular but nonlocal couplings, transitions to IHSS and HSS are observed~\cite{R27}. Here we first consider two layers of network, when oscillators of layer $L_1$ are coupled nonlocally, and oscillators of layer $L_2$ uncoupled, with $R_1=0.08$, and $\epsilon=1.5$, $\lambda_2=0$. The order parameters $A(\lambda_1)$ for both layers in forward and backward directions are shown in Fig.~\ref{f5}a1, b1. The variations of average $x_i (i = 1,2,...,500)$ for both layers $L_1$ and $L_2$ are plotted in Fig.~\ref{f5}a2, b2. It is clear that sudden transitions from oscillatory state to death state and vice versa occur in both forward and backward directions and the death state corresponds to IHSS. We can also see both tipping points or transition points occur at different values of $\lambda_1$ indicating hysteresis in the system. Next, when we increase the range of coupling in $L_1$ with $R_1=0.2$, we observe a sudden transition from oscillatory state to death state in both layers, but there is no transition from death state to oscillatory state in backward direction in both layers (Fig.~\ref{f5}a3, b3). The variations of average $x_i (i = 1,2,...,500)$ for both layers $L_1$ and $L_2$ are plotted in Fig.~\ref{f5}a4, b4. Thus with increasing the range of couplings on layer $L_1$, the transition is changed to tipping with no hysteresis and the steady state is HSS. Due to multiplexing, there is layer to layer synchronization or pairwise in-phase synchronization of nodes between layers, which triggers tipping in both layers.

\begin{figure}
\includegraphics[width=0.5\textwidth]{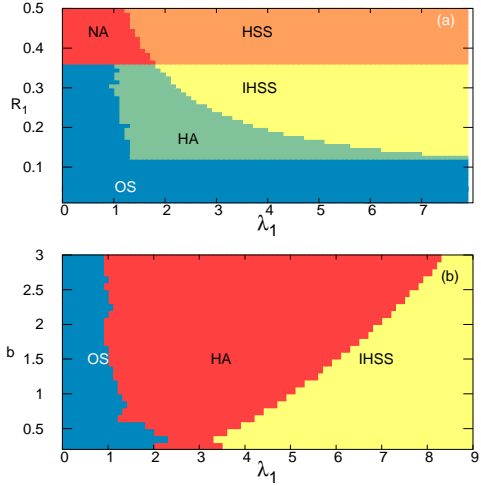}
\caption{ Regions of different emergent states on parameter planes of multiplex network. Here, layer $L_1$ is  with regular and nonlocal connections and layer $L_2$ is uncoupled. {\bf a} $(\lambda_1,R_1)$ at $\lambda_2=0$ {\bf b} $(\lambda_1,b)$ at $\lambda_2=0$ and $R_1=0.1$. The other parameters are chosen as $\epsilon=1.5$, $b=1.0$, and $N=100$.}   
\label{f6}
\end{figure}

Further, we plot the parameter plane $(\lambda_1,R_1)$ at $\lambda_2=0$, for layer $L_1$ in Fig.~\ref{f6}a. In this figure, OS, HA, HSS, IHSS, and NA represent the oscillatory state, hysteresis area, homogeneous steady state,  inhomogeneous steady-state, and no-hysteresis area, respectively.  We can see that when $R_1<0.35$, coupled system has hysteresis,  and the system is stabilized to IHSS. But when $R_1>0.35$, there is no hysteresis in the system, and the system stabilized at coupling dependent HSS. In this case, only forward transition is occurring, which is also sudden, indicating tipping in the absence of multi-stability.  

We also note that when the connections are local (i.e., $R_1=0.01$) there is no transition from oscillatory state to steady-state and for global connection(i.e., $R_1=0.5$), system shows transition only in forward direction. The parameter plane $(\lambda_2,R_2)$ for $L_2$ is similar to that of $L_1$.  We also plot the parameter plane $(\lambda_1,b)$ at $R_1=0.16$, $\lambda_2=0$, for layer $L_1$, which is shown in Fig.~\ref{f6} b. The hysteresis area is seen to increase with the increase in the value of $b$. The parameter planes for layer $L_2$ are qualitatively similar to that of $L_1$.

\begin{figure}
\includegraphics[width=0.75\textwidth]{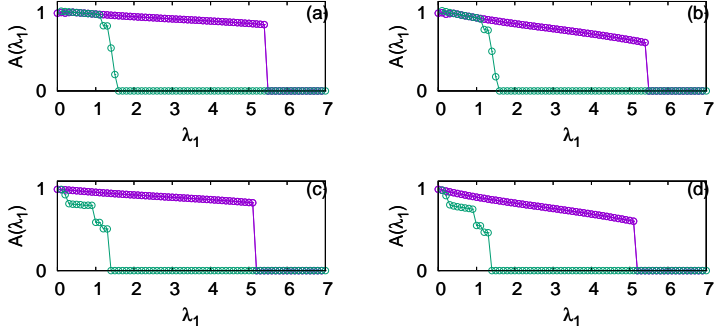}
\caption{Transitions from oscillatory state to death state with hysteresis in multiplex network when both layers have regular and nonlocal connections. The order parameter, $A(\lambda_1)$, under the variation of the coupling strength $\lambda_1$ for layer $L_1$ (left panel) and layer $L_2$ (right panel). {\bf a, b} $R_1 = 0.16$, and $R_2 = 0.04$ and {\bf c, d} $R_1=0.16$ and $R_2=0.16$, with $b=1$, $\epsilon=1.5$, $\lambda_2=0.50$, and  $N=500$.}
\label{f7}   
\end{figure}

Next, we study the case where oscillators of both layers are coupled nonlocally. We first set parameter $R_1=0.16$ $R_2=0.04$ and $\lambda_2=0.5$, and calculate order parameter $A(\lambda_1)$ for both forward and backward directions for both layers $L_1$ and $L_2$. As seen from  Fig.~\ref{f7}a, b, a sudden transition from oscillatory state to death state is observed in forward direction, but in the backward direction, the transition is not sudden, but with hysteresis in the transitions. Also, when we set $R_1=0.16$ $R_2=0.16$ and $\lambda_2=0.5$, such that both layers have identical coupling scheme, we observe the transitions are sudden, which is clear from Fig.~\ref{f7}c, d. 

\begin{figure}[!t]
\includegraphics[width=0.5\textwidth]{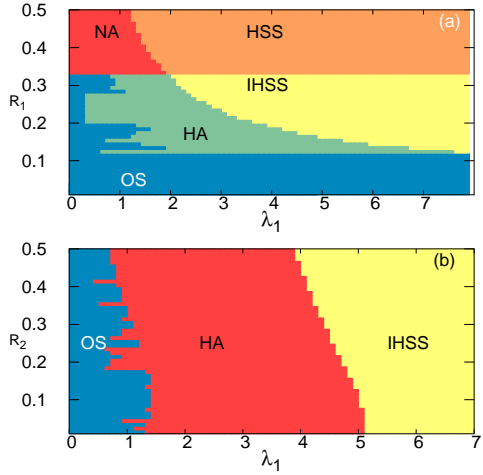}
\caption{Parameter plane of multiplex network with nonlocal connections on both layers. The regions of different emergent states for {\bf a} $(\lambda_1,R_1)$ at $\lambda_2=1.0$ and $R_1=0.10$ {\bf b} $(\lambda_1,R_2)$ at $\lambda_2=0.5$ and $R_1=0.16$. Here $\epsilon=1.5$, $b=1.0$, and $N=100$.}   
\label{f8}
\end{figure}

The parameter plane $(\lambda_1,R_1)$ for layer $L_1$, is plotted at $R_2=0.1$ in Fig.~\ref{f8}a. In this case, system gets stabilized as IHSS when $R_1<0.32$, resulting in transitions with hysteresis.  We also present the parameter plane $(\lambda_1,R_2)$ for $R_1=0.16$ and $\lambda_2=0.5$, which is shown in Fig.~\ref{f8}b. Here also, for a given value of $R_2$, we observe hysteresis and system stabilized as IHSS.

\section{Multiplex network of van der Pol oscillators with mean field couplings}

\begin{figure}
\includegraphics[width=0.65\textwidth]{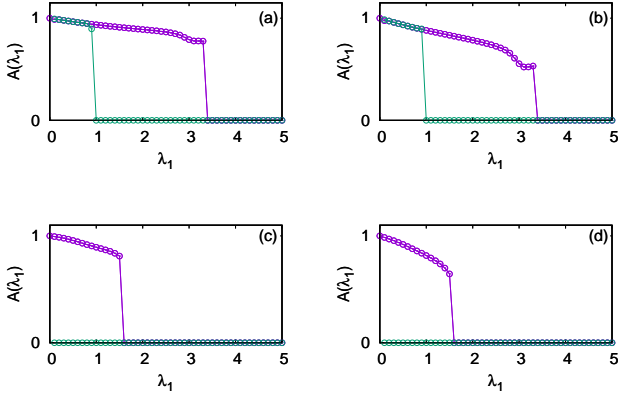}
\caption{ Transitions in multiplex network with mean-field coupling. The order parameter, $A(\lambda_1)$, for transitions in layer $L_1$ (left panel) with mean field couplings under the variation of the coupling strength $\lambda_1$. {\bf a, b} with layer $L_2$ (right panel) uncoupled with $Q_1 = 0.5$, and $Q_2 = 0$ and {\bf c, d} with layer $L_2$  coupled with similar coupling with $Q_1 = 0.5$, $Q_2 = 0.3$, and $\lambda_2 = 1$. $b=2$, $\eta=2$ and $N=100$. The transition to IHSS with hysteresis when layer $L_2$ is uncoupled changes to HSS with no hysteresis when $L_2$ is coupled with the same type of connections as layer $L_1$. }
\label{f9}   
\end{figure}

In order to illustrate the generality of our results above, we consider $N$ identical VdP oscillators coupled via a mean-field diffusive coupling~\cite{R28}. The dynamics of these networks when multiplexed, is given by,

\begin{eqnarray}
\dot{x}_{i,1}&=&y_{i,1}+ \lambda_1(Q_1\bar{x}_{i,1}-x_{i,1})+\eta (x_{i,2}-x_{i,1})  \nonumber\\
\dot{y}_{i,1}&=&b(1-x_{i,1}^2)y_{i,1}-x_{i,1} \nonumber\\
\dot{x}_{i,2}&=&y_{i,2}+\lambda_2(Q_2\bar{x}_{i,2}-x_{i, 2})+\eta(x_{i,1}-x_{i,2})\nonumber\\
\dot{y}_{i,2}&=&b(1-x_{i,2}^2)y_{i,2}-x_{i,2} 
\end{eqnarray}

where parameter $Q_i$,  with $ 0\leq Q_i\leq 1$ (where $i=1,2$), is the intensity of the mean-field and $\lambda_1$ and $\lambda_2$ are the strength of coupling. We start with the case where oscillators of $L_1$ are coupled with each other with mean-field coupling, while oscillators of $L_2$ are uncoupled. We fix $Q_1=0.5$, $b=2$, $\lambda_2=0$, and $\eta=2$, and present the order parameter $A(\lambda_1)$ in both forward and backward directions for both layers, in Fig.~\ref{f9}a, b. We observe sudden transitions with hysteresis. Next, we consider a multiplex network in which oscillators of both layers are coupled among them with mean-field coupling. For $Q_1=0.5$, $Q_2=0.3$, and $\lambda_2=1$ we find sudden transition from oscillatory state to death state in forward direction, with no transition, in the backward direction (Fig.~\ref{f9}c, d). This occurs in both layers, indicating tipping induced from one layer to another layer due to multiplexing. Moreover, the systems in each layer is in complete synchronization among themselves, while they are in pairwise in-phase synchronization between layers.

\begin{figure}
\includegraphics[width=0.6\textwidth]{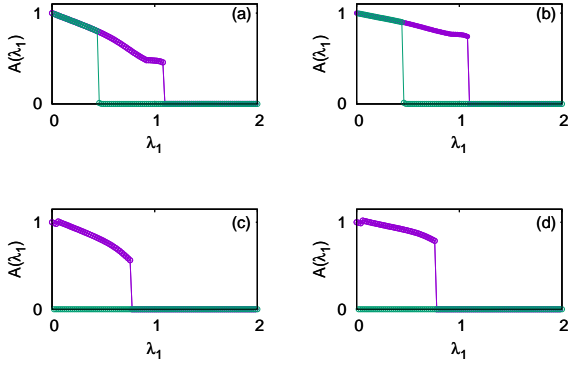}
\caption{Sudden transitions in multiplex network with conjugate couplings. The order parameter, $A(\lambda_1)$, for layer $L_1$ (left panel) with conjugate couplings under the variation of the coupling strength $\lambda_1$. {\bf a, b} when layer $L_2$ (right panel) is uncoupled $\lambda_2 = 0$, showing transition to IHSS with hysteresis and {\bf c, d} layer $L_2$ coupled with similar coupling $\lambda_2 = 0.5$ for transitions to HSS with no hysteresis. We fix $b=2$, $\eta=2$, and $N=100$. }
\label{f10}   
\end{figure}

\section{Multiplex network of van der Pol oscillators with conjugate couplings}
In this section we present our study on multiplex network of VdP oscillators with intralayer coupling between conjugate variables, and the dynamics is then given by equations.
  
  \begin{eqnarray}
\dot{x}_{i,1}&=&y_{i,1}+ \frac{\lambda_1}{N}\sum_{j=1}^{N}(y_{j,1}-x_{i,1}) +\eta (x_{i,2}-x_{i,1}) \nonumber\\
\dot{y}_{i,1}&=&b(1-x_{i,1}^2)y_{i,1}-x_{i,1}++ \frac{\lambda_1}{N}\sum_{j=1}^{N}(x_{j,1}-y_{i,1})\nonumber\\
\dot{x}_{i,2}&=&y_{i,2}+\frac{\lambda_2}{N}\sum_{j=1}^{N}(y_{j,2}-x_{i,2})+\eta(x_{i,1}-x_{i,2})\nonumber\\
\dot{y}_{i,2}&=&b(1-x_{i,2}^2)y_{i,2}-x_{i,2}+\frac{\lambda_2}{N}\sum_{j=1}^{N}(x_{j,2}-y_{i,2}) 
\end{eqnarray}
  
where $i=1,2,3...,N$. As reported earlier,~\cite{R29}, single network with conjugate coupling is shown to have transitions to IHSS and HSS.
We first consider a case where oscillators of layer $L_1$ are coupled globally via conjugate coupling and oscillators of second layer $L_2$ are uncoupled. For this we set $\lambda_2=0$, $\eta=2$, and $b=2$, and calculate order parameter $A(\lambda_1)$ in both forward and backward direction for both layer , which are plotted in Fig.~\ref{f10}a, b respectively. Both layers show a sudden transition from oscillatory state to death state and vice versa. When the oscillators of both layers are globally coupled, with $\lambda_2=0.5$, $\eta=2$ and $b=2$, we observe a sudden transition in the forward direction, (Fig.~\ref{f10} c, d) from oscillatory state to death state, with no backward transition from death state to oscillatory state. In both cases, pair wise in-phase synchronization between layers induces tipping, while layer with conjugate coupling is completely synchronized among its systems.

\section{Multiplex network with heterogeneous topology for layers}

\begin{figure}
\includegraphics[width=0.6\textwidth]{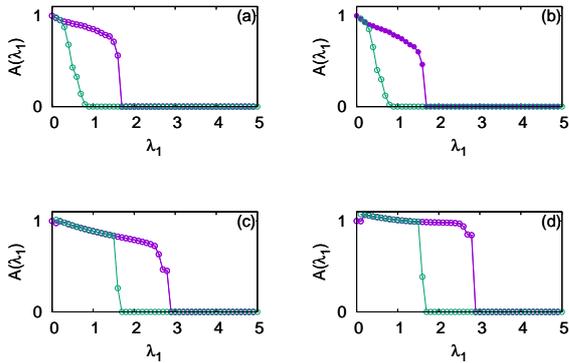}
\caption{ Transitions in multiplex network  with heterogeneous topologies in both layers. The order parameter, $A(\lambda_1)$, under the variation of the coupling strength $\lambda_1$ .{\bf a, b} $L_1$ coupled directly via S-F topology and indirectly via shared environment and $L_2$ coupled locally. Here $< k_1 >= 6$, $R_2 = 0.01$, $b = 1$ and $\lambda_2 = 0.5$. {\bf c, d} $L_1$ coupled via mean field coupling and $L_2$ coupled with nonlocal coupling with $Q_1 = 0.5$, $b = 2$, and $R_2 = 0.5$. Here $\eta=1$ and $N=500$. In both cases transitions occur with hysteresis but in the latter case,  they are sudden in both directions compared to the former type of couplings.}
\label{f11}   
\end{figure}

In this section, we study a multiplex network, where the topology of connections in $L_2$ is different from that in $L_1$. First, we consider a case where oscillators of layer $L_1$ interact directly via S-F topology and indirectly via a shared environment, while oscillators of layer $L_2$ are coupled nonlocally. In this case we set $< k_1 >= 6$ and $R_2 = 0.01$ $b=1$ and $\lambda_2=0.5$, and calculate order parameter $A(\lambda_1)$ in both forward and backward direction for both layers , which are plotted in Fig.~\ref{f11}a, b respectively. Both layers show a sudden transition from oscillatory state to death state in the forward direction, while in the backward direction, the transition is not sudden. Next, we consider a case where oscillators of layer $L_1$ coupled via mean-field diffusive coupling and oscillators of layer $L_2$ coupled nonlocally. We fix $Q1 = 0.5$, $b=2$, and $R_2 = 0.2$, and  the corresponding order parameters in both forward and backward directions for both layers, are shown in Fig.~\ref{f11}c, d respectively. Here in both forward and backward directions, we observe a sudden transition from oscillatory state to death state and vice versa in both layers. In this case with two layers having different topology, we observe pairwise in-phase synchronization of nodes between the layers when multiplexed resulting in tipping in both layers. In general, when both layers have the same type of network topology, the backward transition does not happen, while when both layers have different topology, we observe the backward transition with hysteresis.

\section{Conclusion}
We report the occurrence of sudden transitions or tipping in a collection of systems induced due to multiplexing between two layers of systems. We present mainly the sudden transitions from oscillatory state to death state in both layers of the multiplex network that is found to occur with the same tipping points. Using van der Pol oscillator as the nodal dynamics, we consider three different mechanisms by which sudden transitions in dynamics from oscillations to steady states happen in the network $L_1$ of coupled van der Pol oscillators, coupling with shared environment, mean field coupling and conjugate coupling. In each case, we consider multiplexing with layer $L_2$, with its systems uncoupled, and  coupled with similar and different topologies as that in $L_1$. We consider different topologies like S-F, nonlocal, mean-field, and conjugate couplings for the systems in each layer. We study the possible transitions in each case by computing the order parameter by adiabatic variation of coupling strength  in forward and backward directions, the average values of variables, and parameter plane indicating regions of different emergent dynamical states.

We observe, in general, the nature of transitions depends on the mechanism for tipping in the first layer $L_1$ and the topology of connections in both layers $L_1$ and $L_2$. We find the tipping induced in layer $L_2$ occurs due to pairwise in phase synchronization between the layers due to multiplexing. In general, three types of transitions are observed, sudden transitions in both directions with hysteresis, transitions sudden in forward but not sudden in backward transitions, and sudden transitions in forward direction only with no hysteresis.
When the systems in $L_2$ are uncoupled, in all cases of couplings considered for $L_1$, the tipping on $L_1$ is mirrored exactly on $L_2$. With similar coupling added among systems in the second layer $L_2$ also, the nature of transitions in both layers is modified. Also, the transitions are sudden with hysteresis when the coupled systems stabilize to coupling dependent inhomogeneous steady-states (IHSS), while the transitions are sudden without hysteresis when they stabilize to coupling dependent homogeneous steady-states (HSS). 

When the mechanism for transition in the first layer $L_1$ is due to the presence of a shared environment, with S-F structure, the sudden transition to steady-state on $L_1$ is induced on $L_2$, with hysteresis, both when systems of $L_2$, are coupled and uncoupled. However, with change in average degree introduced on $L_1$, there is tipping, with no hysteresis. For regular networks with nonlocal connections, with an increase in the range of coupling, tipping occurs on both layers, with a sudden transition to HSS.  With mean-field coupling and conjugate coupling as mechanisms for transitions to steady states in $L_1$, with a similar coupling on both layers, we observe sudden transitions to HSS. However, with heterogeneous topology for the layers, the transitions are sudden with hysteresis for mean-field coupling on $L_1$ and nonlocal coupling on $L_2$, while they are sudden in one direction with hysteresis for S-F on $L_1$ and local on $L_2$. In all these cases, the coupling parameters in layer $L_2$ are chosen such that in the absence of multiplexing with $L_1$,  the transitions to steady states do not occur in $L_2$.   This ensures that the critical transitions observed in $L_2$ are induced by those in $L_1$ due to multiplexing.  For this we keep the inter-layer coupling strength above the critical value such that there is pairwise in-phase synchronization between the layers.

Our study is mainly on three types of mechanisms for sudden transitions to steady states on one layer and how it gets induced on another layer when multiplexed, for different network topologies like scale free, regular with nonlocal or global connections. We think our study will have implications that illustrates how tipping can propagate through multiplexed networks, especially for the specific case of meta-communities in ecosystems. In large eco-systems with several patches, tipping can spread across systems due to multiplex mode of migrations between patches in different meta-communities. As in ref~\cite{RN11} , we can consider prey–predator patches communicating in different layers with one topology, but with multiplexing between the two layers of the network that can lead to transitions induced in both layers.

\end{document}